\documentclass[aps,prl,preprint,tightenlines,superscriptaddress,showpacs,byrevtex]{revtex4}

\usepackage{graphicx}
\usepackage{epsfig}
\usepackage{dcolumn}

\newcommand{\lum}{{\cal L}}

\newcommand{\BR}{{\cal B}}

\newcommand{\psp}{\psi^{\prime}}

\newcommand{\jpsi}{J/\psi}

\newcommand{\EE}{e^+e^-}
\newcommand{\MM}{\mu^+\mu^-}
\newcommand{\LL}{\ell^+\ell^-}
\newcommand{\pp}{\pi^+\pi^-}
\newcommand{\kk}{K^+K^-}

\newcommand{\x}{Y(4140)}
\newcommand{\beq}{\begin{equation}}
\newcommand{\eeq}{\end{equation}}
\newcommand{\bitm}{\begin{itemize}}
\newcommand{\eitm}{\end{itemize}}
\begin{document}

%************************************************************
\preprint{} \preprint{\vbox{ \hbox{   }
                        \hbox{Belle Prerpint 2009-24}
                        \hbox{KEK   Preprint 2009-30}
                        \hbox{BIHEP-EP-2009-003}
                        }}
\title{\quad\\[2.0cm]
Evidence for a new resonance and search for the $Y(4140)$ in $\gamma
\gamma \to \phi J/\psi$}

\affiliation{Budker Institute of Nuclear Physics, Novosibirsk}
\affiliation{Faculty of Mathematics and Physics, Charles University,
Prague}
%%%\affiliation{Chiba University, Chiba}
\affiliation{University of Cincinnati, Cincinnati, Ohio 45221}
\affiliation{Department of Physics, Fu Jen Catholic University,
Taipei} \affiliation{Justus-Liebig-Universit\"at Gie\ss{}en,
Gie\ss{}en} \affiliation{The Graduate University for Advanced
Studies, Hayama}
%%%\affiliation{Gyeongsang National University, Chinju}
\affiliation{Hanyang University, Seoul} \affiliation{University of
Hawaii, Honolulu, Hawaii 96822} \affiliation{High Energy Accelerator
Research Organization (KEK), Tsukuba}
%%%\affiliation{Hiroshima Institute of Technology, Hiroshima}
%%%\affiliation{University of Illinois at Urbana-Champaign, Urbana, Illinois 61801}
\affiliation{Institute of High Energy Physics, Chinese Academy of
Sciences, Beijing}
%%%\affiliation{Institute of High Energy Physics, Vienna}
\affiliation{Institute of High Energy Physics, Protvino}
%%%\affiliation{Institute of Mathematical Sciences, Chennai}
\affiliation{INFN - Sezione di Torino, Torino}
\affiliation{Institute for Theoretical and Experimental Physics,
Moscow} \affiliation{J. Stefan Institute, Ljubljana}
\affiliation{Kanagawa University, Yokohama} \affiliation{Institut
f\"ur Experimentelle Kernphysik, Karlsruhe Institut f\"ur
Technologie, Karlsruhe} \affiliation{Korea University, Seoul}
%%%\affiliation{Kyoto University, Kyoto}
\affiliation{Kyungpook National University, Taegu}
\affiliation{\'Ecole Polytechnique F\'ed\'erale de Lausanne (EPFL),
Lausanne} \affiliation{Faculty of Mathematics and Physics,
University of Ljubljana, Ljubljana} \affiliation{University of
Maribor, Maribor} \affiliation{Max-Planck-Institut f\"ur Physik,
M\"unchen} \affiliation{University of Melbourne, School of Physics,
Victoria 3010} \affiliation{Nagoya University, Nagoya}
%%%\affiliation{Nara University of Education, Nara}
\affiliation{Nara Women's University, Nara} \affiliation{National
Central University, Chung-li} \affiliation{National United
University, Miao Li} \affiliation{Department of Physics, National
Taiwan University, Taipei} \affiliation{H. Niewodniczanski Institute
of Nuclear Physics, Krakow} \affiliation{Nippon Dental University,
Niigata} \affiliation{Niigata University, Niigata}
%%%\affiliation{University of Nova Gorica, Nova Gorica}
\affiliation{Novosibirsk State University, Novosibirsk}
\affiliation{Osaka City University, Osaka}
%%%\affiliation{Osaka University, Osaka}
\affiliation{Panjab University, Chandigarh}
%%%\affiliation{Peking University, Beijing}
%%%\affiliation{Princeton University, Princeton, New Jersey 08544}
%%%\affiliation{RIKEN BNL Research Center, Upton, New York 11973}
%%%\affiliation{Saga University, Saga}
\affiliation{University of Science and Technology of China, Hefei}
\affiliation{Seoul National University, Seoul}
%%%\affiliation{Shinshu University, Nagano}
\affiliation{Sungkyunkwan University, Suwon} \affiliation{School of
Physics, University of Sydney, NSW 2006} \affiliation{Tata Institute
of Fundamental Research, Mumbai} \affiliation{Excellence Cluster
Universe, Technische Universit\"at M\"unchen, Garching}
%%%\affiliation{Toho University, Funabashi}
\affiliation{Tohoku Gakuin University, Tagajo} \affiliation{Tohoku
University, Sendai} \affiliation{Department of Physics, University
of Tokyo, Tokyo}
%%%\affiliation{Tokyo Institute of Technology, Tokyo}
%%%\affiliation{Tokyo Metropolitan University, Tokyo}
\affiliation{Tokyo University of Agriculture and Technology, Tokyo}
%%%\affiliation{Toyama National College of Maritime Technology, Toyama}
\affiliation{IPNAS, Virginia Polytechnic Institute and State
University, Blacksburg, Virginia 24061} \affiliation{Yonsei
University, Seoul}
  \author{C.~P.~Shen}\affiliation{Institute of High Energy Physics, Chinese Academy of Sciences, Beijing}\affiliation{University of Hawaii, Honolulu, Hawaii 96822} % Hawaii
  \author{C.~Z.~Yuan}\affiliation{Institute of High Energy Physics, Chinese Academy of Sciences, Beijing} % IHEP
% \author{I.~Adachi}\affiliation{High Energy Accelerator Research Organization (KEK), Tsukuba} % KEK
  \author{H.~Aihara}\affiliation{Department of Physics, University of Tokyo, Tokyo} % Tokyo
  \author{K.~Arinstein}\affiliation{Budker Institute of Nuclear Physics, Novosibirsk}\affiliation{Novosibirsk State University, Novosibirsk} % BINP
% \author{T.~Aso}\affiliation{Toyama National College of Maritime Technology, Toyama} % Toyama
% \author{V.~Aulchenko}\affiliation{Budker Institute of Nuclear Physics, Novosibirsk}\affiliation{Novosibirsk State University, Novosibirsk} % BINP
  \author{T.~Aushev}\affiliation{\'Ecole Polytechnique F\'ed\'erale de Lausanne (EPFL), Lausanne}\affiliation{Institute for Theoretical and Experimental Physics, Moscow} % ITEP
% \author{T.~Aziz}\affiliation{Tata Institute of Fundamental Research, Mumbai} % Tata
% \author{S.~Bahinipati}\affiliation{University of Cincinnati, Cincinnati, Ohio 45221} % Cincinnati
  \author{A.~M.~Bakich}\affiliation{School of Physics, University of Sydney, NSW 2006} % Sydney
  \author{V.~Balagura}\affiliation{Institute for Theoretical and Experimental Physics, Moscow} % ITEP
% \author{Y.~Ban}\affiliation{Peking University, Beijing} % Peking
  \author{E.~Barberio}\affiliation{University of Melbourne, School of Physics, Victoria 3010} % Melbourne
  \author{A.~Bay}\affiliation{\'Ecole Polytechnique F\'ed\'erale de Lausanne (EPFL), Lausanne} % Lausanne
% \author{I.~Bedny}\affiliation{Budker Institute of Nuclear Physics, Novosibirsk}\affiliation{Novosibirsk State University, Novosibirsk} % BINP
  \author{K.~Belous}\affiliation{Institute of High Energy Physics, Protvino} % Protvino
% ...
% BINP
\collaboration{The Belle Collaboration}

\date{\today}

\begin{abstract}

The process $\gamma \gamma \to \phi \jpsi$ is measured for $\phi
\jpsi$ masses between threshold and 5~GeV/${\it c}^2$, using a data
sample of 825~fb$^{-1}$ collected with the Belle detector. A narrow
peak of $8.8^{+4.2}_{-3.2}$ events, with a significance of 3.2
standard deviations including systematic uncertainty, is observed.
The mass and natural width of the structure (named $X(4350)$) are
measured to be $(4350.6^{+4.6}_{-5.1}(\rm{stat})\pm
0.7(\rm{syst}))~\hbox{MeV}/{\it c}^2$ and
$(13^{+18}_{-9}(\rm{stat})\pm 4(\rm{syst}))~\hbox{MeV}$,
respectively.  The product of its two-photon decay width and
branching fraction to $\phi\jpsi$ is $(6.7^{+3.2}_{-2.4}(\rm{stat})
\pm 1.1(\rm{syst}))~\hbox{eV}$ for $J^P=0^+$, or
$(1.5^{+0.7}_{-0.6}(\rm{stat}) \pm 0.3(\rm{syst}))~\hbox{eV}$ for
$J^P=2^+$. No signal for the $Y(4140)\to \phi \jpsi$ structure
reported by the CDF Collaboration in $B\to K^+ \phi \jpsi$ decays is
observed, and limits of $\Gamma_{\gamma \gamma}(\x) \BR(\x\to\phi
\jpsi)<41~\hbox{eV}$ for $J^P=0^+$ or $<6.0~\hbox{eV}$ for $J^P=2^+$
are determined at the 90\% C.L. This disfavors the scenario in which
the $Y(4140)$ is a $D_{s}^{\ast+} {D}_{s}^{\ast-}$ molecule.

\end{abstract}

\pacs{14.40.Rt, 13.25.Gv, 13.66.Bc}

\maketitle

%%%%%%%%%%%%%%%%%%%%%%%%%%%%%%%%%%%%%%%%%%%%%%%%%%%%%%%%%%%%%%%%
%%%%%     Introduction       Part                  %%%%%%%%%%%%%
%%%%%%%%%%%%%%%%%%%%%%%%%%%%%%%%%%%%%%%%%%%%%%%%%%%%%%%%%%%%%%%%
In recent years, many new charmonia or charmonium-like states have
been discovered. These states are not easily accommodated in the
quark-model picture of hadrons~\cite{many}. In this Letter, we
report the first investigation of the $\phi \jpsi$ system produced
in the two-photon process $\gamma \gamma \to \phi \jpsi$ with the
$\jpsi$ decaying into lepton pairs and $\phi \to \kk$, to search for
high mass states with $J^{PC}=0^{++}$ or $2^{++}$, such as the
tetraquark states and molecular states that are predicted by various
models~\cite{tanja, stancu,zhangjr}. We find an unexpected structure
in $\phi \jpsi$ mass near $4350~\hbox{MeV}/{\it c}^2$.

In a related study of $B^+ \to K^+ \phi \jpsi $ decays, the CDF
Collaboration reported evidence of a state called $Y(4140)$ with
mass and width values of $M=(4143.0\pm 2.9(\rm stat)\pm 1.2(\rm
syst))~\hbox{MeV}/{\it c}^2$ and $\Gamma= (11.7^{+8.3}_{-5.0}(\rm
stat)\pm 3.7(\rm syst))~\hbox{MeV}$~\cite{CDF}. The Belle
Collaboration searched for the $Y(4140)$ using the same B mode with
a sample of $772\times 10^6$ $B \bar{B}$ pairs~\cite{lptalk}. No
significant signal was found although the upper limit on the
production rate does not contradict the CDF measurement.

There have been a number of different interpretations proposed for
the $Y(4140)$, including a $D_{s}^{\ast+} {D}_{s}^{\ast-}$
molecule~\cite{tanja, liux, ding, namit,liu3,huang, raphael,molina},
an exotic $1^{-+}$ charmonium hybrid~\cite{namit}, a
$c\bar{c}s\bar{s}$ tetraquark state~\cite{stancu}, or a natural
consequence of the opening of the $\phi\jpsi$ channel~\cite{eef}.
There are arguments against the interpretation of the $Y(4140)$ as a
conventional charmonium state, such as the
$\chi_{c0}^{\prime\prime}$ or
$\chi_{c1}^{\prime\prime}$~\cite{liu2}, or a scalar $D_{s}^{\ast+}
{D}_{s}^{\ast-}$ molecule since QCD sum rules~\cite{wangzg, wangzg2}
predict masses that are inconsistent with the observed value.
Assuming that the $Y(4140)$ is a $D_{s}^{\ast+} {D}_{s}^{\ast-}$
molecule with quantum numbers $J^{PC}=0^{++}$ or $2^{++}$,
Ref.~\cite{tanja} predicts its two-photon width to be of order 1
keV, which can be tested experimentally at Belle.

%%%%%%%%%%%%%%%%%%%%%%%%%%%%%%%%%%%%%%%%%%%%%%%%%%%%%%%%%%%%%%%%
%%%%%  Data sample and MC generator                %%%%%%%%%%%%%
%%%%%%%%%%%%%%%%%%%%%%%%%%%%%%%%%%%%%%%%%%%%%%%%%%%%%%%%%%%%%%%%

This analysis of $\gamma \gamma \to \phi \jpsi$ is based on a
825~fb$^{-1}$ data sample collected with the Belle
detector~\cite{Belle} operating at the KEKB asymmetric-energy $\EE$
collider~\cite{KEKB}. About 90\% of the data were collected at the
$\Upsilon(nS)~(n=1,3,4,5)$ resonances, and about 10\% were taken at
a center-of-mass (C.M.) energy that is 60~MeV below the
$\Upsilon(4S)$ peak.

The detector is described in detail elsewhere~\cite{Belle}. It is a
large-solid-angle magnetic spectrometer that consists of a silicon
vertex detector (SVD), a 50-layer central drift chamber (CDC), an
array of aerogel threshold Cherenkov counters (ACC), a barrel-like
arrangement of time-of-flight scintillation counters (TOF), and an
electromagnetic calorimeter comprised of CsI(Tl) crystals (ECL)
located inside a superconducting solenoid coil that provides a 1.5~T
magnetic field. An iron flux-return located outside of the coil is
instrumented to detect $K_L^0$ mesons and to identify muons (KLM).

We use the program {\sc treps}~\cite{treps} to generate signal Monte
Carlo (MC) events. In this generator, the two-photon luminosity
function is calculated and simulated events are generated at a
specified fixed $\gamma \gamma$ C.M. energy ($W_{\gamma \gamma}$)
using the equivalent photon approximation~\cite{berger}. The
efficiency for detecting $\gamma \gamma \to X \to \phi \jpsi \to \kk
\ell^+ \ell^-$ ($\ell=e$, $\mu$) is determined by assuming
$J^{P}=0^{+}$ or $2^+$ and a zero intrinsic width for the $X$.

%%%%%%%%%%%%%%%%%%%%%%%%%%%%%%%%%%%%%%%%%%%%%%%%%%%%%%%%%%%%%%%%
%%%%%     Event Selection                          %%%%%%%%%%%%%
%%%%%%%%%%%%%%%%%%%%%%%%%%%%%%%%%%%%%%%%%%%%%%%%%%%%%%%%%%%%%%%%

We require four reconstructed charged tracks with zero net charge.
For these tracks, the impact parameters perpendicular to and along
the beam direction with respect to the interaction point are
required to be less than 0.5 and 4~cm, respectively, and the
transverse momentum in the laboratory frame is restricted to be
higher than 0.1~$\hbox{GeV}/c$. For each charged track, information
from different detector subsystems is combined to form a likelihood
$\mathcal{L}_i$ for each particle species~\cite{pid}. Tracks with
$\mathcal{R}_K=\frac{\mathcal{L}_K}{\mathcal{L}_K+\mathcal{L}_\pi}>0.6$,
are identified as kaons with an efficiency of about 97\% for the
tracks of interest; about 0.4\% are misidentified $\pi$
tracks~\cite{pid}. For electron identification, the likelihood ratio
is defined as
$\mathcal{R}_e=\frac{\mathcal{L}_e}{\mathcal{L}_e+\mathcal{L}_x}$,
where $\mathcal{L}_e$ and $\mathcal{L}_x$ are the likelihoods for
electron and non-electron, respectively, determined using the ratio
of the energy deposit in the ECL to the momentum measured in the SVD
and CDC, the shower shape in the ECL, the matching between the
position of charged track trajectory and the cluster position in the
ECL, the hit information from the ACC and the dE/dx information in
the CDC~\cite{EID}. For muon identification, the likelihood ratio is
defined as
$\mathcal{R}_\mu=\frac{\mathcal{L}_\mu}{\mathcal{L}_\mu+\mathcal{L}_\pi+\mathcal{L}_K}$,
where $\mathcal{L}_\mu$, $\mathcal{L}_\pi$ and $\mathcal{L}_K$ are
the likelihoods for muon, pion and kaon hypotheses, respectively,
based on the matching quality and penetration depth of associated
hits in the KLM~\cite{MUID}. For electrons (muons) from $\jpsi$
decay, both of the tracks should have
$\mathcal{R}_e(\mathcal{R}_\mu)>0.1$. The lepton ID efficiency is
about 99\% for $\jpsi\to \EE$ and 90\% for $\jpsi\to \MM$. There are
a few background events due to photon conversions with the
conversion leptons misidentified as kaon candidates in the $\EE$
mode; these are removed by requiring $\mathcal{R}_e < 0.75$ for the
kaon candidates.

\begin{figure}[htbp]
\psfig{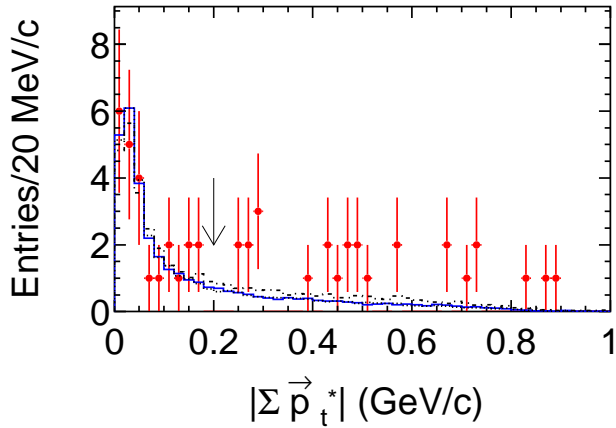} \caption{The magnitude of the
vector sum of $\phi \jpsi$ transverse momenta with respect to the
beam direction in the $\EE$ C.M. frame for the selected $\phi \jpsi$
events. Points with error bars are data. The dot-dashed,  solid, and
dotted histograms are MC simulations for $\gamma \gamma \to \phi
\jpsi$ with the $\phi\jpsi$ mass fixed at 4.20, 4.35, and
4.50~GeV/$c^2$, respectively (normalized to the number of events
with $|\sum {\vec P}_t^{\ast}|<0.2$~GeV/$c$). The arrow shows the
position of the $|\sum {\vec P}_t^{\ast}|$ requirement.} \label{xpt}
\end{figure}

The magnitude of the vector sum of the four tracks' transverse
momenta in the C.M. frame, $|\sum {\vec P}_t^{\ast}|$, which
approximates the transverse momentum of the two-photon-collision
system, is required to be less than $0.2~\hbox{GeV}/c$ in order to
reduce backgrounds from non-two-photon processes and
two-photon-processes with extra particles other than $\phi$ and
$\jpsi$ in the final states. Figure~\ref{xpt} shows the $|\sum {\vec
P}_t^{\ast}|$ distributions from the data and a MC signal simulation
before the $|\sum {\vec P}_t^{\ast}|$ requirement. Here the $\kk$
and $\ell^+ \ell^-$ invariant masses are required to be within the
$\phi$ and $\jpsi$ signal regions, respectively.

\begin{figure}[htbp]
\psfig{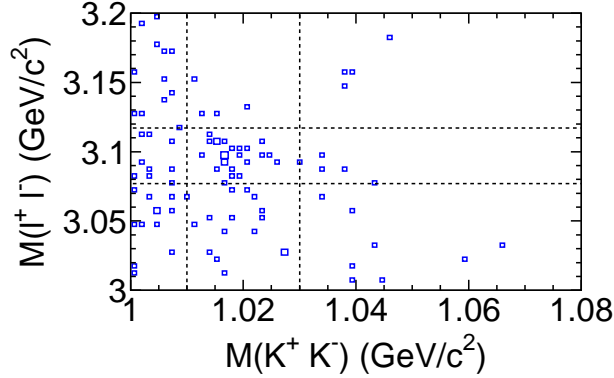} \caption{A scatter plot of
$M(\ell^+\ell^-)$ versus $M(\kk)$ for the selected $\kk \ell^+
\ell^-$ events. The size of the boxes is proportional to the number
of events. } \label{scatter}
\end{figure}

A scatter plot of $M(\ell^+\ell^-)$ versus $M(\kk)$ for the selected
$\kk \ell^+ \ell^-$ events is shown in Fig.~\ref{scatter}, where we
can see clear $\jpsi$ and $\phi$ signals. A partial correction for
final state radiation and bremsstrahlung energy loss is performed by
including the four-momentum of every photon detected within a
50~mrad cone around the electron and positron direction in the $\EE$
invariant mass calculation. We define a $\jpsi$ signal region as
$3.077~\hbox{GeV}/c^2<m_{\ell^+\ell^-}<3.117~\hbox{GeV}/c^2$ (the
mass resolution is about 10~MeV/${\it c}^2$), and $\jpsi$ mass
sidebands as
$3.0~\hbox{GeV}/c^2<m_{\ell^+\ell^-}<3.06~\hbox{GeV}/c^2$ or
$3.14~\hbox{GeV}/c^2<m_{\ell^+\ell^-}<3.20~\hbox{GeV}/c^2$. We also
define a $\phi$ signal region as
$1.01~\hbox{GeV}/c^2<m_{\kk}<1.03~\hbox{GeV}/c^2$ (the full width at
half maximum (FWHM) of the $\phi$ signal is 5.9~MeV/$c^2$), and
$\phi$ mass sidebands as
$1.00~\hbox{GeV}/c^2<m_{\kk}<1.01~\hbox{GeV}/c^2$ or
$1.03~\hbox{GeV}/c^2<m_{\kk}<1.08~\hbox{GeV}/c^2$.

%%%%%%%%%%%%%%%%%%%%%%%%%%%%%%%%%%%%%%%%%%%%%%%%%%%%%%%%%%%%%%%%
%%%%%   KKLL mass spectrum and FIT                 %%%%%%%%%%%%%
%%%%%%%%%%%%%%%%%%%%%%%%%%%%%%%%%%%%%%%%%%%%%%%%%%%%%%%%%%%%%%%%

\begin{figure}[htbp]
\psfig{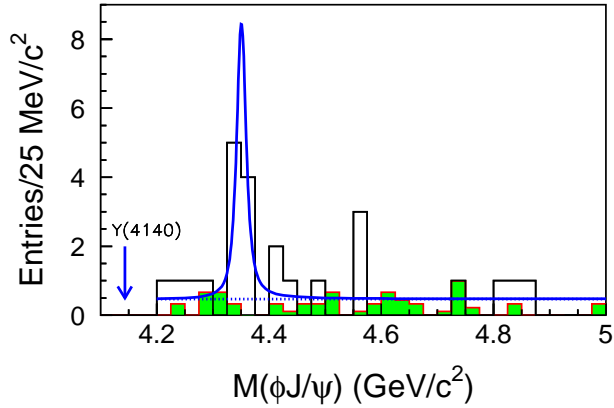}\caption{The $\phi \jpsi$ invariant
mass distribution of the final candidate events. The open histogram
shows the experimental data. The fit to the $\phi \jpsi$ invariant
mass distribution from 4.2 to 5.0 GeV/$c^2$ is described in the
text. The solid curve is the best fit, the dashed curve is the
background, and the shaded histogram is from normalized $\phi$ and
$\jpsi$ mass sidebands. The arrow shows the expected position of the
$\x$.} \label{mkkjpsi-fit2}
\end{figure}

Figure~\ref{mkkjpsi-fit2} shows the $\phi \jpsi$ invariant mass
distribution~\cite{mass}, together with the background estimated
from the normalized $\jpsi$ and $\phi$ mass sidebands. No $\x$
signal is evident. Assuming that there is no background within the
$\x$ mass region and the number of signal events follows a Poisson
distribution with a uniform prior probability density function, a
Bayesian upper limit on the number of the $Y(4140)$ signal events is
estimated to be 2.3 at the 90\% C.L.~\cite{PDG}. However, there is a
clear enhancement at 4.35~GeV/$c^2$, where the background level
estimated from the normalized $\jpsi$ and $\phi$ mass sidebands is
very low. Other possible backgrounds that are not included in the
sidebands, such as $\gamma\gamma \to \phi \jpsi +X$ and $\EE \to
\phi \jpsi +X$ where $X$ may indicate one or more particles, and
$\gamma\gamma\to \phi \jpsi$ with the $\jpsi$ and $\phi$ decaying
into final states other than lepton pairs and $\kk$, are found to be
very small after applying all of the event selection criteria.

In order to obtain resonance parameters for the structure at
$4.35~\hbox{GeV}/c^2$, an unbinned extended maximum likelihood
method is applied to the $\phi \jpsi$ mass spectrum in
Fig.~\ref{mkkjpsi-fit2}. The distribution is fitted in the range 4.2
to 5.0~GeV/$c^2$ with an acceptance-corrected Breit-Wigner (BW)
function convoluted with a double Gaussian resolution function as
the signal shape and a constant term as the background shape. The
shape of the double Gaussian resolution function is obtained from MC
simulation.

From the fit, we obtain $8.8^{+4.2}_{-3.2}$ signal events, with a
mass $m=(4350.6^{+4.6}_{-5.1})$~MeV/${\it c}^2$ and a width
$\Gamma=(13^{+18}_{-9})$~MeV. The statistical significance of this
structure is estimated to be $3.9\sigma$, from the difference of the
logarithmic likelihoods, $-2\ln(L_0/L_{\rm max})=21$, taking the
difference in the number of degrees of freedom ($\Delta
\hbox{ndf}=3$) in the fits into account, where $L_0$ and $L_{\rm
max}$ are the likelihoods of the fits with and without a resonance
component, respectively. In the following we refer to it as the
$X(4350)$. The significance of the signal decreases to $3.2\sigma$
if a linear function is used to model the background shape in the
fit.

We use an ensemble of simulated experiments to estimate the
probability that background fluctuations alone would produce signals
as significant as that seen in the data. We generate $\phi\jpsi$
mass spectra based on a uniform distribution alone with 24 events,
the same as observed in data, and search for the most significant
fluctuation in each spectrum in the mass range from 4.2 to
5.0~GeV/$c^2$, with widths in a range between 3~MeV (half of the
resolution) and 130~MeV (ten times the observed width). From these
spectra we obtain the distribution for $-2\ln(L_0/L_{\rm max})$ in
pure background samples, and compare it with the signal in the data.
We performed a total of 0.5 million simulations and found 65 trials
with a $-2\ln(L_0/L_{\rm max})$ value greater than or equal to the
value obtained in the data. The resulting $p$ value is $1.3\times
10^{-4}$, corresponding to a significance of $3.8\sigma$.
Generating events in a wider $\phi J/\psi$ mass range, or fitting with
different width range would change the resulting significance, but
the dependence is weak for a signal as narrow as the $X(4350)$.

%%%%%%%%%%%%%%%%%%%%%%%%%%%%%%%%%%%%%%%%%%%%%%%%%%%%%%%%%%%%%%%%
%%%%%   The products of two-photon width and Brs   %%%%%%%%%%%%%
%%%%%%%%%%%%%%%%%%%%%%%%%%%%%%%%%%%%%%%%%%%%%%%%%%%%%%%%%%%%%%%%

The product of the two-photon decay width and branching fraction is
obtained using the formula: $\Gamma_{\gamma \gamma}(R)\BR(R\to
\hbox{final~state})=N/[(2J+1)\epsilon\, {\cal K} \lum_{\rm int}]$,
where $N$ is the number of observed events, $\epsilon$ is the
efficiency, $J$ is the spin of the resonance, and $\lum_{\rm int}$
is the integrated luminosity. ${\cal K}$ is a factor that is
calculated from the two-photon luminosity function ${\cal
L}_{\gamma\gamma}(M_R)$ for a resonance with mass $M_R$ using the
relation: ${\cal K}=4\pi^2{\cal L}_{\gamma\gamma}(M_R)/M_R^2$, which
is valid when the resonance width is small compared to its mass
(widths are smaller than 1\% of the masses in the $Y(4140)$ and
$X(4350)$ cases). The ${\cal K}$ parameter is calculated to be
$0.46~\hbox{fb}/\hbox{eV}$ and $0.36~\hbox{fb}/\hbox{eV}$ for the
$\x$ and the $X(4350)$, respectively, using {\sc
treps}~\cite{treps}. The efficiencies are 0.30\% and 0.41\% for
$J^P=0^+$ and $2^+$, respectively, at $4.143~\hbox{GeV}/c^2$, and
7.90\% and 6.98\% for $J^P=0^+$ and $2^+$ respectively, at
$4.35~\hbox{GeV}/c^2$. From the above values, we obtain
$\Gamma_{\gamma\gamma}(\x)\BR(\x\to\phi \jpsi)<36~\hbox{eV}$ for
$J^P=0^+$, or $<5.3~\hbox{eV}$ for $J^P=2^+$, at the 90\% C.L., and
$\Gamma_{\gamma \gamma}(X(4350))\BR(X(4350)\to \phi
\jpsi)=(6.7^{+3.2}_{-2.4})~\hbox{eV}$ for $J^P=0^+$, or
$(1.5^{+0.7}_{-0.6})~\hbox{eV}$ for $J^P=2^+$, where the errors are
statistical only.

%%%%%%%%%%%%%%%%%%%%%%%%%%%%%%%%%%%%%%%%%%%%%%%%%%%%%%%%%%%%%%%%
%%%%%     Systematic Errors       Part            %%%%%%%%%%%%%
%%%%%%%%%%%%%%%%%%%%%%%%%%%%%%%%%%%%%%%%%%%%%%%%%%%%%%%%%%%%%%%%
There are several sources of systematic errors for the measurements
of the products of the two-photon decay width and branching
fractions. The particle identification uncertainties are 1.2\%/kaon
and 0.8\%/lepton. The uncertainty in the tracking efficiency for
tracks from $\jpsi$ decays is 1\% per track, while that for kaon
tracks from $\phi$ decays range from 2.4\% to 1\% per track as the
average transverse momentum increases from 0.15 to 0.3~GeV/$c$. The
efficiency uncertainties associated with the $\jpsi$ and $\phi$ mass
requirements are determined from the studies of the very pure $\EE
\to \psp \to \pi^+ \pi^- \jpsi$~\cite{isr-ppjpsi} and $\EE \to \phi
\pp$~\cite{isr-phipp} event samples. The detection efficiencies for
$\jpsi$ and $\phi$ mesons are lower than those inferred from the MC
simulations by $(2.5\pm0.4)\%$ and $(2.0\pm0.5)\%$ relatively,
respectively. We take 0.96 as the efficiency correction factor, and
0.7\% is included in the systematic error due to the $\jpsi$ and
$\phi$ mass requirements. The statistical errors in the MC samples
are 2.3\% and 0.9\% for $\x \to \phi \jpsi$ and $X(4350)\to \phi
\jpsi$, respectively. The accuracy of the two-photon luminosity
function calculated by the {\sc treps} generator is estimated to be
about 5\% including the error from neglecting radiative corrections
(2\%), the uncertainty from the form factor effect (2\%), and the
uncertainty in the total integrated luminosity (1.4\%)~\cite{treps}.
The trigger efficiency for four charged track events is rather high
because of the redundancy of the Belle first level multi-track
trigger. According to the MC simulation, the trigger and
preselection efficiency for the final state has little dependence on
the $\phi \jpsi$ invariant mass, with an uncertainty that is smaller
than 5\%. From Ref.~\cite{PDG}, the uncertainty in the world average
values for $\BR(\phi\to \kk)$ is 1.2\% and that for $\BR(\jpsi\to
\LL)=\BR(\jpsi\to \EE)+\BR(\jpsi\to \MM)$ is 1\% where we have added
the errors of the $\EE$ and $\MM$ modes linearly. The uncertainty in
the yield of $X(4350)$ signal events due to the $\phi \jpsi$ mass
spectrum fit is estimated to be 15\% by varying: the order of the
background polynomial (14\%), resonance parameterization (0.5\%) and
the $\phi \jpsi$ mass resolution (0.9\%).
%These errors are summarized in
%Table~\ref{err_psp}.
Assuming that all the sources are independent and adding all
uncertainties in quadrature, we obtain the total systematic errors
on $\Gamma_{\gamma \gamma}(\x) \BR(\x \to \phi \jpsi)$ and
$\Gamma_{\gamma \gamma}(X(4350)) \BR(X(4350) \to \phi \jpsi)$ to be
12\% and 17\%, respectively.

For the systematic errors in the $X(4350)$ mass and width, the
uncertainties in the mass resolution (0.1~MeV/${\it c}^2$ and
0.9~MeV), the parameterization of the resonance (0~MeV/${\it c}^2$
and 0.3~MeV) and the background shape (0.7~MeV/${\it c}^2$ and
3.9~MeV) are considered. Assuming that all the sources are
independent and adding them in quadrature, we obtain the total
systematic errors on the $X(4350)$ mass and width to be
0.7~MeV/${\it c}^2$ and 4.1~MeV, respectively.

%%%%%%%%%%%%%%%%%%%%%%%%%%%%%%%%%%%%%%%%%%%%%%%%%%%%%%%%%%%%%%%%
%%%%%     Summary       Part                       %%%%%%%%%%%%%
%%%%%%%%%%%%%%%%%%%%%%%%%%%%%%%%%%%%%%%%%%%%%%%%%%%%%%%%%%%%%%%%

In summary, we report results of the first search for $Y(4140)\to
\phi\jpsi$ in the two-photon process $\gamma \gamma \to \phi \jpsi$.
No $\x$ signal is observed, and upper limits on the product of the
two-photon decay width and branching fraction of $\x \to \phi \jpsi$
are established to be $\Gamma_{\gamma \gamma}(\x) \BR(\x\to\phi
\jpsi)<41~\hbox{eV}$ for $J^P=0^+$, or $<6.0~\hbox{eV}$ for
$J^P=2^+$ at the 90\% C.L. In the determination of the
$\Gamma_{\gamma \gamma}(\x) \BR(\x \to \phi \jpsi)$ upper limits,
the efficiencies have been lowered by a factor of $1-\sigma_{sys}$
to obtain a conservative estimate, where $\sigma_{sys}$ is the total
relative systematic error. The upper limit on $\Gamma_{\gamma
\gamma}(\x) \BR(\x\to\phi \jpsi)$ from this experiment is lower than
the prediction of $(176^{+137}_{-93})$~eV for $J^{PC}=0^{++}$,
$(189^{+147}_{-100})$~eV for $J^{PC}=2^{++}$ (calculated by us using
the values in Ref.~\cite{tanja} and total width of the $Y(4140)$
from CDF~\cite{CDF}). This disfavors the scenario in which the
$Y(4140)$ is a $D_{s}^{\ast+} {D}_{s}^{\ast-}$ molecule with
$J^{PC}=0^{++}$ or $2^{++}$.

We find evidence for an unexpected new narrow structure at
$4.35~\hbox{GeV}/c^2$ in the $\phi\jpsi$ mass spectrum with a
significance of 3.2 standard deviations including systematic
uncertainty. If this structure is interpreted as a resonance, its
mass and width are $(4350.6^{+4.6}_{-5.1}(\rm{stat})\pm
0.7(\rm{syst}))~\hbox{MeV}/{\it c}^2$ and
$(13^{+18}_{-9}(\rm{stat})\pm 4(\rm{syst}))~\hbox{MeV}$,
respectively. The product of its two-photon decay width and
branching fraction to $\phi \jpsi$ is measured to be $\Gamma_{\gamma
\gamma}(X(4350)) \BR(X(4350)\to\phi
\jpsi)=(6.7^{+3.2}_{-2.4}(\rm{stat}) \pm 1.1(\rm{syst}))~\hbox{eV}$
for $J^P=0^+$, or $(1.5^{+0.7}_{-0.6}(\rm{stat}) \pm
0.3(\rm{syst}))~\hbox{eV}$ for $J^P=2^+$. We note that the mass of
this structure is consistent with the predicted values of a
$c\bar{c}s\bar{s}$ tetraquark state with $J^{PC}=2^{++}$ in
Ref.~\cite{stancu} and a $D^{\ast+}_s {D}^{\ast-}_{s0}$ molecular
state in Ref.~\cite{zhangjr}. In a recent paper~\cite{liu4}, the
possibility that the $X(4350)$ could be an excited $P$-wave
charmonium state, $\chi_{c2}''$, was also discussed.

%%%%%%%%%%%%%%%%%%%%%%%%%%%%%%%%%%%%%%%%%%%%%%%%%%%%%%%%%%%%%%%%
%%%%%    acknowledgments       Part                %%%%%%%%%%%%%
%%%%%%%%%%%%%%%%%%%%%%%%%%%%%%%%%%%%%%%%%%%%%%%%%%%%%%%%%%%%%%%%

We thank the KEKB group for excellent operation of the
accelerator, the KEK cryogenics group for efficient solenoid
operations, and the KEK computer group and the NII for valuable
computing and SINET3 network support. We acknowledge support from
MEXT, JSPS and Nagoya's TLPRC (Japan); ARC and DIISR (Australia);
NSFC (China); DST (India); MEST, KOSEF, KRF (Korea); MNiSW
(Poland); MES and RFAAE (Russia); ARRS (Slovenia); SNSF
(Switzerland); NSC and MOE (Taiwan); and DOE (USA).


\begin{thebibliography}{**}
\bibitem{mass} In this Letter, $m_{\kk\LL}-m_{\LL}-m_{\kk}+m_{\jpsi}+m_{\phi}$
is used instead of the invariant mass of the four final-state
particles to improve the mass resolution. Here $m_{\jpsi}$ and
$m_\phi$ are the nominal masses of $\jpsi$ and $\phi$~\cite{PDG}.
\end{thebibliography}
\end{document}